\documentstyle[11pt,aaspp]{article}

\begin{document}
\title{A Calculation of the Full Neutrino Phase Space in Cold+Hot
Dark Matter Models}
\author{Chung-Pei Ma and Edmund Bertschinger}
\affil{Center for Theoretical Physics, Laboratory for Nuclear Science
and Department of Physics, Massachusetts Institute of Technology,
Cambridge, MA 02139}
\begin{abstract}
This paper presents a general-relativistic $N$-body technique for
evolving the phase space distribution of massive neutrinos in  linear
perturbation theory.  The method provides a much more accurate sampling
of the neutrino phase space for the HDM initial conditions of $N$-body
simulations in a cold+hot dark matter (CDM+HDM) universe than previous
work.  Instead of directly sampling the phase space at the end of the
linear era, we first compute the evolution of the metric perturbations
by numerically integrating the coupled, linearized Einstein, Boltzmann,
and fluid equations for all particle species (CDM, baryons, photons,
massless neutrinos, and massive neutrinos).  (Details of this
calculation are discussed in a separate paper.)  We then sample the
phase space shortly after neutrino decoupling at redshift $z=10^9$
when the distribution is Fermi-Dirac.  To follow the trajectory of each
neutrino, we subsequently integrate the geodesic equations for each
neutrino in the perturbed background spacetime from $z = 10^9$ to
$z=13.55$, using the linearized metric found in the previous calculation
to eliminate discreteness noise.  The positions and momenta resulting
from this integration represent a fair sample of the full neutrino phase
space and can be used as HDM initial conditions for $N$-body simulations
of nonlinear structure evolution in CDM+HDM models.  A total of $\sim21$
million neutrino particles are used in a 100 Mpc box, with
$\Omega_{\rm cdm}=0.65$, $\Omega_{\rm hdm}=0.30$, $\Omega_{\rm baryon}=
0.05$, and Hubble constant $H_0=50$ km s$^{-1}$ Mpc$^{-1}$.
We find that correlations develop in the neutrino densities and momenta
which are absent when only the zeroth-order Fermi-Dirac distribution
is considered.

\end{abstract}

\section{Introduction}
As it has became increasingly difficult to explain cosmological
observations in the context of the standard cold dark matter (CDM)
model, the cold+hot dark matter (CDM+HDM) models have emerged as one
of the promising alternatives that require only moderate modifications
of the CDM model.  The excess small-scale power relative to the
large-scale power in the standard CDM model (Davis et al. 1985, 1992a;
Gelb, Gradw\"ohl, \& Frieman 1993; Gelb \& Bertschinger 1993) motivates
the addition of a hot component of massive neutrinos to the total mass
density of the universe.  The free streaming of the neutrinos suppresses
the growth of small-scale perturbations while leaving the growth of
large-scale perturbations unimpeded, and may therefore alleviate some
of the problems of the standard CDM model.  Several recent linear
calculations (Schaefer, Shafi, \& Stecker 1989; van Dalen \& Schaefer
1992; Taylor \& Rowan-Robinson 1992; Holtzman \& Primack 1993) and
$N$-body simulations (Davis, Summers, \& Schlegel 1992b; Klypin et al.
1993) have found a better match of observations with the CDM+HDM
models than with the CDM models, although a fair comparison between the
models and the galactic scale data such as the epoch of galaxy formation
and galaxy pairwise velocities awaits results from higher resolution
$N$-body simulations in a large volume.  Most workers (including ourselves)
have assumed that the mass density fraction contributed by HDM is
$\Omega_{\rm hdm} \sim 0.3$ for an $\Omega_{\rm total}=1$ universe,
corresponding to a neutrino mass of $m_\nu \sim 7$ eV, although
Pogosyan \& Starobinsky (1993) favor $0.17 \le \Omega_{\rm hdm} \le 0.28$
for $H_0=50$ km s$^{-1}$ Mpc$^{-1}$.

The introduction of HDM into the theory brings about one complication
due to the different behavior of the neutrinos and CDM.  In the linear
regime CDM behaves as a pressureless perfect fluid, but the neutrinos
can be appropriately described only by their full phase space distribution
obeying the Boltzmann equation.  None of the earlier studies of HDM
models of which we are aware has taken account of the full phase space
information of the neutrinos.

In the particle-particle/particle-mesh (P$^3$M) simulation performed
by Davis et al. (1992b), the HDM particles were placed initially on a
grid without perturbations at $1+z=20$, based on the argument that
the neutrino Jeans length at this redshift is comparable to their
simulation box size of 14 Mpc.  The initial conditions for the CDM
particles were generated with the Zel'dovich approximation from the
{\it pure} CDM spectrum and scaled to the normalization $\sigma_8 \sim
0.45$.  To simulate the thermal motion of the particles, a velocity
drawn randomly from the Fermi-Dirac distribution was given to each
HDM particle.  The cosmological parameters $\Omega_{\rm cdm}=0.7$,
$\Omega_{\rm hdm}=0.3$, and $H_0=50$ km s$^{-1}$ Mpc$^{-1}$ were used,
with $32^3$ CDM and $32^3$ HDM particles.
%

In the particle-mesh (PM) simulations of Klypin et al. (1993), the
initial conditions for CDM and HDM were generated with the Zel'dovich
method from individual CDM and HDM power spectra, which differ
significantly on scales smaller than the neutrino free-streaming
distance since the growth of perturbations in HDM is suppressed.
A thermal velocity drawn from the Fermi-Dirac distribution was
added to each HDM particle.  The simulations were performed in 14, 50,
and 200 Mpc boxes with $128^3$ CDM and $6\times 128^3$ HDM particles
starting at redshift $1+z =15$.  The parameters $\Omega_{\rm
cdm}=0.6$, $\Omega_{\rm hdm}=0.3$, $\Omega_{\rm baryon}=0.1$, and
$H_0 = 50$ km s$^{-1}$ Mpc$^{-1}$ were used.

In both groups' simulations, the initial neutrino momenta were
drawn from the Fermi-Dirac distribution independently of the neutrino
positions.  In general, however, the neutrino phase space distribution
is a complicated function of positions, momenta (or velocities), and
time, with the Fermi-Dirac distribution being only the zeroth-order
term.  Velocity-position correlations in the neutrinos can arise from
perturbations to the Fermi-Dirac distribution.
Although the actual HDM correlations are initially small
in the linear regime, they can play an important role in the nonlinear
stage of evolution, and in the linear theory should be treated as being
of the same order as all other perturbations.

In this paper, we obtain the neutrino initial conditions from the full
phase space distribution in the linear theory of gravitational
perturbations.  Since the full phase space distribution depends on 6
canonical variables, it is numerically impractical to sample
the full neutrino phase space at the start of $N$-body simulations.
We resolve this difficulty by sampling the neutrino phase space with
neutrino particles at a very early time of $z \sim 10^9$ when the spatial
distribution of neutrinos is nearly uniform.  At this time the phase
space distribution is Fermi-Dirac to a very good approximation, and
the neutrinos can be placed on a grid with momenta drawn from the
Fermi-Dirac distribution.  Since the neutrinos have already decoupled
from other species ($z_{\nu, {\rm dec}} \sim 10^{10}$), their
trajectories simply follow geodesics in the perturbed Robertson-Walker
spacetime.

Our strategy will be to first calculate the metric perturbations by
integrating the coupled, linearized Einstein, Boltzmann, and fluid
equations that govern the evolution of the metric and density
perturbations of all particle species (CDM, photons, baryons, massless
neutrinos, and massive neutrinos).  Then we integrate the linearized
geodesic equations for each neutrino from $z \sim 10^9$ until
$z = 13.55$, after which we will switch to a fully nonlinear Newtonian
integration.  The high-redshift approach can be described as a
``general-relativistic cosmological $N$-body integration'' valid in
the linear theory.  It differs from the conventional $N$-body technique
in that the gravitational forces are precomputed from the metric
perturbations of the background spacetime rather than directly from
the particles.  The configuration of neutrinos at $z = 13.55$ will
then represent a fair sample of the full phase space distribution at
that redshift, and can be used directly as the HDM initial conditions
for subsequent $N$-body simulations.

We leave the discussion of the first stage of our calculation on the
Einstein, Boltzmann, and fluid equations to a separate paper (Ma and
Bertschinger 1993).  In the present paper we focus on the geodesic
integration assuming the metric perturbations have been computed.  We
find the conformal Newtonian gauge a very convenient choice for this
part of the calculation.  We derive the linearized geodesic equations
in this gauge in Section 2 and discuss the integration method in
Section 3.  We report the integration results in Section 4 where
we show the effect of the perturbations in the neutrino
phase space on the correlation between the HDM momenta and the
density contrast.  Section 5 includes a summary and a discussion of
work in progress.

\section{Geodesic Equations in Conformal Newtonian Gauge}
\label{section:geodesic}
Although many calculations of the general-relativistic
linear perturbation theory have been carried out in the
synchronous gauge, we find it most convenient to compute the
trajectories of the neutrinos in the conformal Newtonian gauge
(Mukhanov, Feldman, \& Brandenberger 1992).
The conformal Newtonian gauge has the advantage that spurious coordinate
singularities do not arise, and the geodesic equations have simple forms
which are easy to integrate.  The metric in the conformal Newtonian gauge
is given by
\begin{equation}
    ds^2 = a^2(\tau)\left[ -(1+2\psi)d\tau^2 +
        (1-2\phi)\gamma_{ij}dx^i dx^j \right] \ ,
\label{conformal}
\end{equation}
where the scalar potentials $\psi$ and $\phi$ characterize
the perturbations about a flat Robertson-Walker spacetime.
We use Cartesian coordinates so that the 3-metric of $\tau=\hbox{constant}$
hypersurfaces is $\gamma_{ij}=\delta_{ij}$.
It should be emphasized that $\phi$ and $\psi$ describe only the
scalar mode of the metric perturbations.  We do not consider the
vector and the tensor modes in this paper.

The geodesic equations for a neutrino of mass $m_\nu$ can be derived by
minimizing the action
\begin{equation}
        S = \int d\tau L = -m_\nu \int (-ds^2)^{1/2}\ ,
\end{equation}
where $L$ is the Lagrangian and the metric in the conformal Newtonian
gauge is given by Eq.~(\ref{conformal}).
To linear order in the potentials, the Lagrangian is
\begin{equation}
        L = -m_\nu a \sqrt{1-u^2} \left( 1 +
        {\psi+u^2\phi \over 1-u^2} \right) \ ,
\end{equation}
where $\vec{u} = d\vec{x}/d\tau$ is the coordinate velocity and
$u^2=\gamma_{ij}u^iu^j$.  The conjugate momentum $q_i$ is given by
\begin{equation}
      q_i \equiv {\partial L \over \partial u^i}
        = {m_\nu a\gamma_{ij}u^j \over \sqrt{1-u^2}} \left(
        1-2\phi-{\psi+u^2\phi \over 1-u^2}
        \right) \ ,
\label{q}
\end{equation}
which can be inverted to give, to first order in $\psi$ and $\phi$,
\begin{equation}
   u^i = {dx^i \over d\tau} = {\gamma^{ij}q_j \over \epsilon(q,\tau)}
    \left\{ 1 + \psi(\vec{x},\tau) + \left[2-{q^2 \over
        \epsilon^2(q,\tau)}\right] \phi(\vec{x},\tau) \right\}
\label{geox}
\end{equation}
with $\epsilon(q,\tau)=\sqrt{q^2+m^2_\nu a^2}$.
The Euler-Lagrange equation of motion gives
\begin{equation}
        { dq_i \over d\tau} = -{m_\nu a \over \sqrt{1-u^2}}
        \left( \partial_i{\psi} + u^2 \partial_i{\phi}
        \right) \ .
\end{equation}
Replacing $\vec{u}$ on the right-hand side with Eq.~(\ref{geox}),
we obtain
\begin{equation}
   {dq_i \over d\tau} = -\epsilon(q,\tau) \left[
        \partial_i \psi(\vec{x},\tau) + {q^2 \over \epsilon^2(q,\tau)}
        \partial_i \phi(\vec{x},\tau) \right]\,.
\label{geoq}
\end{equation}
Eqs.~(\ref{geox}) and (\ref{geoq}) give the linearized geodesic
equations for a particle moving in the perturbed spacetime characterized
by scalar metric perturbations $\psi$ and $\phi$.  In the weak-field,
nonrelativistic ($q^2\ll\epsilon^2$) limit they reduce to the standard
Newtonian equations.

\section{Integration of Geodesic Equations}
To sample the neutrino phase space as accurately as possible, we
integrate the geodesic equations (\ref{geox}) and (\ref{geoq}) for
$10\times 128^3$ ($\sim$ 21 million) massive neutrino particles.  A
cubic simulation box with sides 100 Mpc is used.  The cosmological
parameters are taken to be $\Omega_{\rm cdm}=0.65$, $\Omega_{\rm
hdm}=0.3$, $\Omega_{\rm baryon}=0.05$, and $H_0 = 50$ km s$^{-1}$
Mpc$^{-1}$.  We start the integration shortly after neutrino
decoupling at redshift $z\sim 10^9$ when perturbations in the neutrino
density and momenta can be safely ignored.  The neutrinos are
placed initially on a $128^3$ grid, 10 per grid point, with the
neutrino momenta drawn randomly from the Fermi-Dirac distribution
\begin{equation}
        f(\vec q\,)\,d^3q \propto {d^3q\over e^{qc / kT_{\nu,0}} + 1} \ ,
\label{fd}
\end{equation}
where $T_{\nu\,,0}=(4/11)^{1/3}\,T_{\gamma\,,0}$ is the neutrino
temperature today with $T_{\gamma\,,0}=2.735$ K.  We performed test
runs with the same set of momenta but randomly generated initial
positions and found no statistically significant difference at the
end of the integration depending on whether the neutrinos were
initially placed at random or on a grid.

We also tested the momentum pairing scheme used by Klypin et al.
(1993) in their initial conditions.  For every momentum drawn from the
Fermi-Dirac distribution, they assigned an equal but opposite momentum
to a second neutrino at the same grid point to preserve the local
center of momentum.  We performed two test runs, with the initial neutrino
momenta drawn randomly in one run and paired up in opposite directions
with the same magnitude in the other run.  We found no statistically
significant difference in the power spectrum at the end of the
geodesic integration.  We thus adopted the simpler scheme without
pairing.

We integrated the geodesic equations from conformal time
$\tau_i=3\times 10^{-4}$ Mpc ($z \sim 10^9$) to $\tau_f=3\times 10^3$
Mpc ($z = 13.55$), using 701 time steps with stepsize
$\Delta(\log_{10}\tau) = 0.01$.  The initial $\tau_i$ is chosen so
that the largest $k$ in the simulation box is well outside the horizon
($k\tau\ll1$) at the onset of the integration.  The integration was
stopped when the fluctuations were still in the linear regime.  We
used a leap-frog integration scheme in which the positions and momenta
were advanced half a timestep out of phase to give a second-order accuracy
in timestep size.

The evolution of the metric perturbations $\psi$ and $\phi$ in
Eqs.~(\ref{geox}) and (\ref{geoq}) were precomputed from the coupled,
linearized Einstein, Boltzmann, and fluid equations for all particle
species including massive neutrinos (Ma and Bertschinger 1993).  The
resulting transfer functions were saved on a grid of 41 $k$- and 701
$\tau$- values.  For the geodesic integration the initial $\psi$ and
$\phi$ were generated as Gaussian random variables in $k$-space with
the scale-invariant power spectrum $P_\psi\propto k^{-3}$ predicted
by the simplest inflationary cosmology models.  For later times, the
Fourier components of $\psi$ and $\phi$ simply scale according to our
linear theory computation.  We normalized the amplitude to the COBE rms
quadrupole fluctuation $Q_{\rm rms-PS} = 14\times 10^{-6}$ K (Seljak \&
Bertschinger 1993; Wright et al. 1992; Smoot et al. 1992) assuming the
Sachs-Wolfe formula for a scale-invariant spectrum and $T_{0\,,\gamma}
=2.735$ K:
\begin{equation}
	{Q_{\rm rms-PS}^2 \over T_{0\,,\gamma}^2}
	= {5\over 108}\,\left[4\pi k^3P_\psi(k,\tau_{\rm rec})
	\right]_{k\to0}
	\ .
\label{SW}
\end{equation}
The gradients of $\psi$ and $\phi$ in Eqs.~(\ref{geox}) and (\ref{geoq})
were first computed on a grid in $k$-space and then Fourier transformed
to a grid in real space.  The second-order Triangular-Shaped Cloud (TSC)
interpolation scheme was then used to interpolate the gradients from
the grid to the particle positions (see Ma 1993 for more details).

Test runs were performed with $N_{\rm grid}=32^3$ using $ N_{\rm
part}=10\times 32^3$, $40\times 32^3$, and $80\times 32^3$, and $N_{\rm
grid}=64^3$ using $N_{\rm part}=10\times 64^3$ and $40\times 64^3$.
(The first factor in $N_{\rm part}$ gives the number of samples of the
momentum space at each initial position.)  We tested the accuracy of
the time integration using 351 and 701 timesteps respectively and
found little difference in the final positions and velocities,
indicating that 701 timesteps are sufficient.  We also generated
realizations of the potentials and the initial neutrino momenta using
three different random number generators.  No correlations in the
random numbers were detected.  Our large production run had $N_{\rm
grid}=128^3$ and $N_{\rm part}=10\times 128^3$ ($\sim$ 21 million).
The geodesic integration required a total of $\sim 1.5$ Gbytes of memory
and $\sim 140$ CPU hours on the Convex C3880 supercomputer at the
National Center for Supercomputing Applications.

\section{Numerical Results}

An image of an intermediate output (timestep 351) from one of the
$N_{\rm part}=10\times 32^3$ test runs is shown in Fig.~\ref{hc.32.3}.
The corresponding redshift is $z \sim 4.9 \times 10^5$.  Each side in
the figure is 100 Mpc comoving, and the particles in the simulation
box have been projected onto the $x-y$ plane.  At the starting $z \sim
10^9$, the particles were placed on a $32^3$ grid, 10 per grid point,
and were given momenta drawn randomly from the Fermi-Dirac distribution.
In this figure, one sees that the neutrinos have begun to spread out
from the grid points.  In fact, the size of each ``ball'' is
approximately the comoving horizon distance $c\tau \sim 0.95$ Mpc at
this moment since the neutrinos are still relativistic.

Fig.~\ref{hc.128.7} shows the same projection in a 100 Mpc box of the
last output (timestep 701) from the $N_{\rm part} = 10\times 128^3$
run.  The corresponding redshift is $z = 13.55$.  As one can see,
small perturbations are developing in the otherwise uniform
distribution of the neutrinos.  We present quantitative analyses of
this output below.

To check the integration results, we computed the HDM power spectrum
from the final output of the $N_{\rm part} = 10\times 128^3$ run and
made comparison with the prediction from the linear theory.  The
density field $\delta$ was first computed on a $128^3$ spatial grid
from the positions of the neutrinos by the TSC interpolation scheme,
and then Fourier transformed into $k$-space.  We calculated the power
per $\ln k$ at a given $k$, $4\pi k^3 P(k)$, by taking a spherical
shell of radius $k$ and thickness $\Delta k$ centered at the origin of
the $k$-grid and averaging the contribution to $4\pi k^3 P(k)$ from
the grid points that lie within the shell.  The shot noise due to the
finite number of particles was subtracted and the TSC window was
deconvolved.  The result is shown in Fig.~\ref{fig:p128} and is
compared with the linear theory predictions for the CDM and HDM power
spectra.  The agreement provides an important check of the accuracy of
our geodesic integration code.  From smaller simulations we conclude
that the deviations from the ensemble-average power HDM spectrum are
due to sampling fluctuations.

The output shown in Fig.~\ref{hc.128.7} is used as the initial
conditions for the HDM particles in our $N$-body simulations of
structure formation in this CDM+HDM model.  To compare our initial
conditions to those of others, we recall that the initial positions
and velocities of the particles in $N$-body simulations are
conventionally generated from the power spectrum using the
Zel'dovich (1970) approximation.  In this procedure, the positions
of the particles are displaced from a regular grid:
\begin{equation}
        \vec{x}(\tau) = \vec{x}_0 + \vec{\epsilon}\,(\vec{x}_0,\tau)\ ,
\end{equation}
where $\vec{x}_0$ gives the position of the grid and
$\vec{\epsilon}\,(\vec{x}_0,\tau)$ is the displacement field.  The
displacements are computed from the density perturbation field by
solving $\vec{\nabla}\cdot\vec{\epsilon} = - \delta \,$.  For small
displacements, $\vec{\epsilon}\,(\vec{x}_0,\tau)$ is approximated by
$D_+(\tau)\,\vec{\epsilon}\,(\vec{x}_0)$, where $D_+(\tau)$ denotes
the growth factor of the perturbations, and the velocities of particles
are given by
\begin{equation}
 \vec{v} \equiv {d\vec{x}\over d\ln a} = {d\ln D_+\over d\ln a}
        \,\vec{\epsilon}\,(\vec{x}_0,\tau)\,.
\end{equation}

For the standard CDM model with $\Omega =1$, the growth factor in the
matter-dominated era is equal to the expansion factor, and $f(\Omega)
\equiv d\ln D_+/d\ln a =1$. The growth rate, however, does not behave
so simply in models such as the CDM+HDM models where more than one
particle species contributes to $\Omega$.  This is illustrated by the
power spectra shown in Fig.~\ref{fig:powspec} for the standard CDM model
and our CDM+HDM model.  The growth rate of CDM in the CDM+HDM model
matches the growth rate in the standard CDM model only at small $k$.
At large $k$ where neutrino free-streaming is important, we have
$\delta_{\rm hdm} \ll \delta_{\rm cdm}$, and $f(\Omega,k) = d\ln D_+
/d\ln a < 1$ and is $k$-dependent.  We calculated $f(\Omega,k)$ for CDM
at $z=13.55$ in our CDM+HDM model from the output of the linear theory
integration; the result is shown in Fig.~\ref{fomega}.  In the limit
$\delta_{\rm cdm} = \delta_{\rm baryon}$ and $\delta_{\rm hdm}=0$, the
growth rate can be computed analytically to be $f = (\sqrt{1+24
\Omega_{\rm c}}-1)/4$ where $\Omega_{\rm c}=\Omega_{\rm cdm+baryon}$
(Bond, Efstathiou, \& Silk 1980).  For our parameters, $f=0.805$.
As one can see in Fig.~\ref{fomega}, $f$ is indeed approaching this
value at high $k$.

If one did not take into consideration the $k$-dependence of the
growth rate and instead used $f(\Omega,k)=1$ to obtain the CDM initial
velocities in CDM+HDM models, one would give the CDM particles
excessive initial velocities on small scales, leading to earlier
gravitational collapses in the simulations.  We estimated this effect
in the linear theory on scales below the free-streaming distance.  We
find that using $f=1$ instead of $f=0.805$ gives an initial amplitude in
the linear growing mode that is too large by a factor $1.093$.  Since
galaxies form later in CDM+HDM models than in the standard CDM model
due to neutrino free-streaming, the epoch of structure formation is
one of the crucial factors that will determine the fate of the model.
By overestimating $f(\Omega,k)$, one underestimates the severity of
the problem with late galaxy formation in CDM+HDM models.
Klypin et al. (1993) set $f=1$ for the CDM.  However, this error was
cancelled by an opposite effect (Primack, private communication): they
used the baryon transfer function for the CDM.  The baryonic perturbations
are smaller than the CDM by up to 15\%.  As a result, the CDM particles
were given less power, which counteracted their excessive velocities
so that the two errors essentially cancelled.

In the simulations by Davis et al. (1992b), the initial HDM momenta
were drawn randomly from the Fermi-Dirac distribution.  In Klypin et
al. (1993), this thermal velocity was added to the velocity arising
from the Zel'dovich approximation for each HDM particle.  At their
starting $z \sim 15$, the thermal component was about a factor of 4
larger.  Neither group included the actual correlations between
neutrino positions and momenta that develop through the Boltzmann
equation in the linear regime.  To incorporate these first-order
effects, we retained the full phase space information by sampling the
phase space at $z\sim 10^9$ with 21 million neutrino particles and
following their trajectories in the perturbed background spacetime
until low redshifts.

To estimate the importance of these neutrino phase space perturbations,
we mimicked the approach of Klypin et al. to generate an ``equivalent''
set of initial positions and momenta at $z = 13.55$ for $10\times
128^3$ neutrinos, using the same realization for $\delta$ as for
$\phi$ and $\psi$ in our geodesic integrations.  A randomly-drawn
thermal velocity was also added to the Zel'dovich velocity for each
neutrino particle.

We calculated $\delta_{\rm hdm}$ from the particle positions and
examined the correlation between the rms neutrino velocities and
$\delta_{\rm hdm}$ in the two cases.  If the neutrinos obeyed the
zeroth-order Fermi-Dirac distribution, the neutrino velocities should
be uncorrelated with $\delta_{\rm hdm}\,$; any correlation would
indicate deviations from the Fermi-Dirac distribution.  Our results
are plotted in Fig.~\ref{vdel}.  We see that correlations in the
velocities and the density perturbations have developed in the
Boltzmann integration case between $z\sim 10^9$ and $z = 13.55$.
The more clustered neutrinos appear to have higher rms velocities
and therefore higher temperature, possibly resulting from the increase
in the kinetic energy when the neutrinos fall into the CDM potential
wells.  This correlation is absent when the initial conditions are
generated with the conventional Zel'dovich approach because the dominant
thermal contribution to the neutrino velocities is drawn from the
zeroth-order Fermi-Dirac distribution with constant temperature.

To test whether gravitational infall is responsible for the density-velocity
correlation, we also computed for each neutrino particle the velocity
components parallel and perpendicular to the gravitational acceleration
$\vec{g} = -\vec{\nabla}\phi$ at the location of the neutrino:
$v_\parallel = \vec{v}\cdot\hat{g}$ and $v_\perp = \sqrt{v^2 -
v_\parallel^2}$.  Then we calculated the conditional rms $v_\parallel$
and $v_\perp$ for a given $\delta_{\rm hdm}$.  The results are shown in
Fig.~\ref{vpara}, where the solid curve represents $\langle
v_\parallel^2 \rangle^{1/2}$ and the dashed curve represents
$\langle v_\perp^2 \rangle^{1/2}/\sqrt{2}$.  The component along the
gravitational acceleration is larger than the orthogonal components by
$\sim 2\%$.  Thus, the velocity-density correlation is not simply due
to a uniform gravitational infall.  This is because the velocity
dispersion (temperature) of the neutrinos and not just the bulk (fluid)
velocity is higher in the denser regions.

For comparison, Fig.~\ref{vcdm} shows the rms velocity versus the
density for our CDM particles at redshift $z = 13.55$.  The
positions and the velocities were generated using the method described
earlier.  As one can see from Figs.~\ref{vdel} and \ref{vcdm}, CDM is
more clustered than HDM (the range of $\delta$ is larger), and the rms
velocities of the CDM particles are $\sim 30$ km s$^{-1}$ compared to
$\sim 95 - 105$ km s$^{-1}$ for the HDM particles.  The CDM velocities
show no significant correlation with $\delta_{\rm cdm}$.  No correlation
is expected in linear theory.

Fig.~\ref{contour} is a contour plot of neutrino density in the velocity
component-$\delta_{\rm hdm}$ plane from the last output of the $N_{\rm
part} = 10\times 128^3$ geodesic integrations.  The rapid decline with
$v$ is due to the (approximately) Fermi-Dirac distribution (with rms
$v\sim55$ km s$^{-1}$) and the decline with $\vert\delta\vert$ is due to
the Gaussian distribution of the potential.  However, the contours are
asymmetric about $\delta_{\rm hdm} = 0$, showing a positive correlation
in the velocities and the density perturbations in the HDM component.
One sees that the overdense regions contain hotter neutrinos than the
underdense regions.

\section{Conclusion and Work in Progress}
Motivated by the CDM+HDM models, we have presented a
general-relativistic $N$-body technique that provides an accurate
sampling of the full neutrino phase space at all times when the linear
perturbation theory is valid.  Although the evolution of the neutrino
phase space distribution can be solved from the Boltzmann equation, we
know of no practical scheme for computing and sampling the final
distribution except for the Monte Carlo method we have employed.
In this method we first compute the metric perturbations about a
Robertson-Walker spacetime by integrating the coupled, linearized
Boltzmann, Einstein, and fluid equations for all particle species,
including the massive neutrinos.  Then we sample the massive neutrino
phase space right after neutrino decoupling at $z\sim 10^9$ when the
distribution is Fermi-Dirac to a very good approximation.  We subsequently
integrate the linearized geodesic equations for individual neutrinos to
obtain their trajectories in the perturbed background spacetime described
by the metric perturbations found in the previous calculation.  This
technique is valid only in the linear regime.  It differs from the
conventional $N$-body simulation method in that the gravitational forces
are precomputed from the metric perturbations of the background spacetime
using continuum linear theory rather than from the particles directly.
The resulting neutrino positions and velocities can be used as the HDM
initial conditions for subsequent $N$-body simulations of the nonlinear
evolution of structures in the CDM+HDM models.

The same method could be used to generate initial conditions for the pure
HDM model.  Although these would differ from what previous workers have
assumed, because of the very large damping of small-scale fluctuations
we are doubtful that there would be significant differences in one's
conclusions about the model.

We are currently performing a high resolution particle-particle
particle-mesh (P$^3$M) $N$-body simulation of the nonlinear
evolution of the density perturbations, using the positions and
velocities from the last output of our large geodesic integration as
the HDM initial conditions.  The initial conditions for the CDM
particles were generated from the CDM power spectrum with a modified
form of the Zel'dovich approximation taking into account the
wavenumber-dependence of the growth rate $f=d\ln D_+/d\ln a$.
A total of $10\times 128^3$ HDM and $128^3$ CDM particles are used
in a 100 Mpc comoving box starting at $z = 13.55$.  If computer time
permits, we will also perform an ``equivalent'' simulation with the
same parameters but with the Zel'dovich initial conditions adopted
by other groups.  Our simulation box will be large enough to include
most of the important long-wavelength power absent in smaller boxes,
and the P$^3$M force calculation will give us much higher resolution
than particle-mesh (PM) simulations.  In addition, we will be able
to make a fair comparison of the two different treatments of the
initial conditions and examine the importance of correlations in the
neutrino phase space.  Until that time it would be premature for us
judge the merits of the approximate methods used by previous workers.

\acknowledgments

This work was supported by NSF grant AST90-01762 and DOE grant
DE-AC02-76ER03069.  Supercomputer time
was provided by the National Center for Supercomputing Applications.
We appreciate the advice and comments of Alan Guth and Jim Frederic.

\begin{figure}
   \vspace{5.5truein}
   \caption{Neutrinos in a 100 Mpc box at $z \sim 4.9\times 10^5$
        (timestep 351, $c\tau=0.95$ Mpc) from one of the
	$N_{\rm part}=10\times 32^3$ test runs.  Projection of the box
	onto the $x-y$ plane is shown.  Each ball is a horizon-radius
	shell of relativistic neutrinos expanding away from their original
	grid positions.}
    \plotfiddle{hc.32.35.ps}{0pt}{0}{100}{100}{-305}{-120}
    \label{hc.32.3}
\end{figure}
\begin{figure}
   \vspace{5.5truein}
   \caption{Neutrinos in a 100 Mpc box at $z = 13.55$
        (timestep 701) from the $N_{\rm part}=10\times 128^3$
        run.  Projection of the box onto the $x-y$ plane is shown.
        The grayscale represents the projected density.}
    \plotfiddle{hc.128.7.ps}{0pt}{0}{100}{100}{-305}{-100}
    \label{hc.128.7}
\end{figure}
\begin{figure}
   \vspace{3.5truein}
    \caption{The power per $\ln k$ (fill circles) computed from
	$10\times 128^3$ neutrino particles at the end of the geodesic
	integration ($z = 13.55$).  The CDM and HDM power spectra
	computed from the linear theory are shown (dotted curves) for
	comparison.}
    \plotfiddle{p128.ps}{0pt}{0}{85}{85}{-260}{-220}
    \label{fig:p128}
\end{figure}
\begin{figure}
   \vspace{4.8truein}
    \caption{The dashed curves are the power per $\ln k$ of CDM and HDM
        in a hybrid model with $\Omega_{\rm cdm}=0.65$, $\Omega_{\rm
        hdm}=0.3$ and $\Omega_{\rm baryon}=0.05$ at redshift $z
        = 13.55$.  For comparison, the solid curve represents the
        CDM power in a CDM model with $\Omega_{\rm cdm}=0.95$ and
        $\Omega_{\rm baryon}=0.05$ at the same redshift.}
    \plotfiddle{powspec.ps}{0pt}{0}{85}{85}{-270}{-200}
    \label{fig:powspec}
\end{figure}
\begin{figure}[t]
   \vspace{3.8truein}
    \caption{The growth rate $f(\Omega,k)=d\ln D_+/d\ln a$ of CDM
        at $z=13.55$ in the CDM+HDM model with $\Omega_{\rm cdm}=0.65$,
        $\Omega_{\rm hdm}=0.3$ and $\Omega_{\rm baryon}=0.05$.}
    \plotfiddle{fomega.ps}{0pt}{0}{75}{75}{-230}{-190}
    \label{fomega}
\end{figure}
\begin{figure}
   \vspace{3.5truein}
    \caption{The rms neutrino velocities $\langle
        (d|\vec{x}|/d\ln a)^2 \rangle^{1/2}$ versus
        the density perturbation $\delta$ at $z = 13.55$.  The solid
        line is computed from $10\times 128^3$ neutrinos at the end of
        the geodesic equation integration.  The dashed line represents
        an ``equivalent'' set of initial conditions generated from the
        Zel'dovich method.  The density bins have size $\Delta\delta_
	{\rm hdm}=0.02$ and the rms velocities of the particles in each
	bin are shown.}
    \plotfiddle{vdel.ps}{0pt}{0}{85}{85}{-260}{-200}
    \label{vdel}
\end{figure}
\begin{figure}
   \vspace{3.7truein}
   \caption{The rms neutrino velocity components versus
        $\delta_{\rm hdm}$.  The solid curve represents
        $\langle v_\parallel^2 \rangle^{1/2}$, where $v_\parallel$ is
        the velocity component parallel to the gravitational
        acceleration; the dashed curve represents the perpendicular
        velocity component $\langle v_\perp^2
        \rangle^{1/2}/\protect\sqrt{2}$.}
    \plotfiddle{vpara.ps}{0pt}{0}{85}{85}{-260}{-220}
   \label{vpara}
\end{figure}
\begin{figure}
   \vspace{3.7truein}
   \caption{The rms CDM velocities versus $\delta_{\rm cdm}$.
   The velocities are entirely parallel to the direction of gravity.}
    \plotfiddle{vcdm.ps}{0pt}{0}{85}{85}{-260}{-240}
    \label{vcdm}
\end{figure}
\begin{figure}
   \vspace{3.7truein}
    \caption{Contour plot of constant particle number in the
        neutrino velocity component-$\delta_{\rm hdm}$ plane.
        The absolute values of all three velocity components are
        shown.  The five contours from bottom up correspond to 10$^5$,
        10$^4$, 1000, 100 and 10 neutrinos per pixel.  Each pixel has
	a width $\Delta\delta=0.02$ and a height $\Delta(dx/d\ln a)=10$
	km s$^{-1}$.}
    \plotfiddle{contour.ps}{0pt}{0}{85}{85}{-260}{-220}
   \label{contour}
\end{figure}

\end{document}